\documentclass[aps,prl,reprint,superscriptaddress]{revtex4-1} 
\usepackage{graphicx}  % needed for figures
\usepackage{dcolumn}   % needed for some tables
\usepackage{physics}
\usepackage{amssymb}
\usepackage{natbib}
\usepackage{color}

\begin{document}

\title{Voltage-Controlled Superconducting Quantum Bus}
\author{L.~Casparis}
\affiliation{Center for Quantum Devices, Station Q Copenhagen, Niels Bohr Institute, University of Copenhagen, Copenhagen, Denmark}
\author{N.~J.~Pearson}
\affiliation{Center for Quantum Devices, Station Q Copenhagen, Niels Bohr Institute, University of Copenhagen, Copenhagen, Denmark}
\affiliation{Theoretische Physik, ETH Z{\"u}rich, Z{\"u}rich, Switzerland}
\author{A.~Kringh\o{}j}
\affiliation{Center for Quantum Devices, Station Q Copenhagen, Niels Bohr Institute, University of Copenhagen, Copenhagen, Denmark}
\author{T.~W.~Larsen}
\affiliation{Center for Quantum Devices, Station Q Copenhagen, Niels Bohr Institute, University of Copenhagen, Copenhagen, Denmark}
\author{F.~Kuemmeth}
\affiliation{Center for Quantum Devices, Station Q Copenhagen, Niels Bohr Institute, University of Copenhagen, Copenhagen, Denmark}
\author{J.~Nyg\r{a}rd}
\affiliation{Center for Quantum Devices, Station Q Copenhagen, Niels Bohr Institute, University of Copenhagen, Copenhagen, Denmark}
\affiliation{Nano-Science Center, Niels Bohr Institute, University of Copenhagen, Copenhagen, Denmark}
\author{P.~Krogstrup}
\affiliation{Center for Quantum Devices, Station Q Copenhagen, Niels Bohr Institute, University of Copenhagen, Copenhagen, Denmark}
\author{K.~D.~Petersson}
\affiliation{Center for Quantum Devices, Station Q Copenhagen, Niels Bohr Institute, University of Copenhagen, Copenhagen, Denmark}
\author{C.~M.~Marcus}
\affiliation{Center for Quantum Devices, Station Q Copenhagen, Niels Bohr Institute, University of Copenhagen, Copenhagen, Denmark}

\begin{abstract}
We demonstrate the ability of an epitaxial semiconductor-superconductor nanowire to serve as a field-effect switch to tune a superconducting cavity. Two superconducting gatemon qubits are coupled to the cavity, which acts as a quantum bus. Using a gate voltage to control the superconducting switch yields up to a factor of 8 change in qubit-qubit coupling between the on and off states without detrimental effect on qubit coherence. High-bandwidth operation of the coupling switch on nanosecond timescales degrades qubit coherence. 
\end{abstract}

%\pacs{03.67.Lx, 81.07.Gf, 85.25.Cp}
\maketitle

%%Introduction
A significant challenge to scaling any quantum processor architecture is controlling interactions between qubits for multiqubit operations. Couplings between superconducting qubits are commonly controlled by detuning their transition frequencies~\cite{dicarlo_2009,kelly_2014}. In this way, instead of changing the qubit-qubit coupling constant, the effective coupling can be suppressed by making the qubit energies non-degenerate~\cite{majer_2007}. As superconducting qubits scale to larger networks, however, the increasingly crowded spectrum of qubit transition frequencies will make this approach prohibitively difficult. Increased frequency crowding makes residual couplings harder to suppress. Moreover, rearranging qubit frequencies, as is likely required during multiqubit operations, can lead to state leakage, as described by Landau-Zener physics~\cite{landau_1932,zener_1932,stuckelberg_1932,majorana_1932}. For reasonable device parameters this results in leakage of several percent~\cite{quintana_2013}. On-chip switchable coupling is desirable, since there is a trade-off between fast two qubit gates and avoiding state leakage.

Tunable coupling schemes have been realized for nearest-neighbour-coupled flux-tunable qubits~\cite{bialczak_2011,chen_2014}, as well as fixed frequency qubits~\cite{mckay_2016}. These schemes allow qubits to be isolated for certain operations, for instance frequency retuning or single qubit rotations, while still enabling fast two qubit gates. A tunable superconducting microwave resonator has also been proposed for selective qubit coupling~\cite{wallquist_2006}. Such an approach has the advantage that a superconducting resonator, acting as a quantum bus, can mediate long range interactions between superconducting qubits, and also allows increased connectivity between qubits~\cite{wallraff_2004,majer_2007,sillanpaa_2007,dicarlo_2009}. Experimentally, flux control of resonators has been demonstrated~\cite{sandberg_2008,palacios_2008} and used to couple superconducting qubits to spin ensembles~\cite{kubo_2011}.

While superconducting qubit circuits often use on-chip current lines to generate fluxes for control, the recently introduced gatemon superconducting qubit~\cite{delange_2015,larsen_2015} is based on a voltage tunable semiconductor Josephson junction (JJ). Gatemons therefore allow for operation using voltages, which can be readily screened to minimize crosstalk and are compatible with semiconductor-based cryogenic control logic~\cite{ward_2013,al-taie_2013,hornibrook_2014}. The advantage of voltage-controlled operation of semiconductor JJs suggests wider applications in a variety of superconducting circuits, such as superconducting field effect transistors (SFETs)~\cite{Qi_2018}.

\begin{figure}[!b]\vspace{-4mm}%aspect ratio: 241/157 -> [(150 / ) + 20 words] = 118 words
    \includegraphics[width=1\columnwidth]{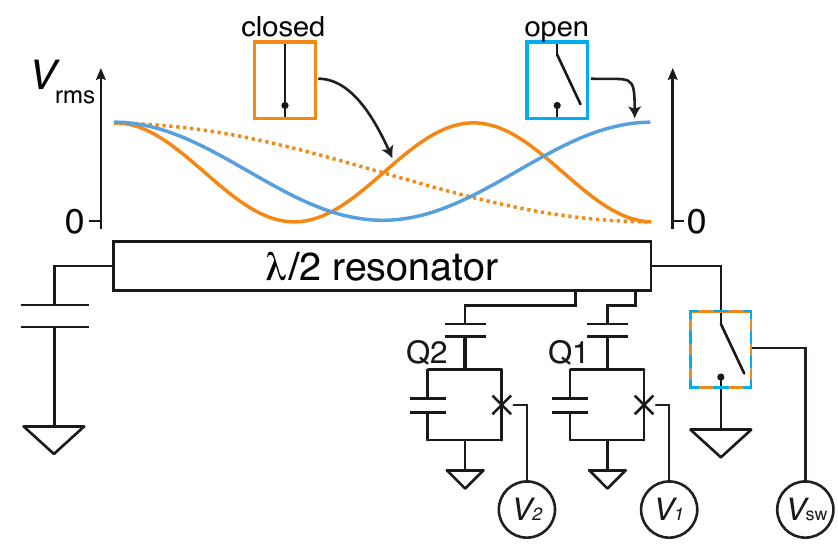}
    \caption{Schematic of the voltage controlled superconducting quantum bus. Two gatemon qubits are capacitively coupled to a $\lambda$/2 resonator. One end of the resonator can be grounded through a voltage controlled superconducting switch. Depending on the switch position being open (blue), or closed (orange), the rms voltage along the resonator length is changed, modifying the coupling between qubits by effectively turning the $\lambda$/2 resonator (blue) into a $\lambda$/4 resonator (orange). The dashed and solid orange lines represent the first and second modes of the $\lambda$/4 resonator respectively.
    }
    \label{device}
\end{figure}

Here, we implement a voltage controllable superconducting resonator -- a tunable quantum bus -- which is strongly coupled to two gatemon qubits. The bus is terminated by an SFET acting as a switch that allows \textit{in situ} control of the resonator frequency and qubit-resonator coupling. We demonstrate that the coupling between the two gatemons can be switched between `on' and `off' states by controlling the SFET with on/off coupling ratios up to $\sim$8. We also show that when the coupling is turned off, the frequency of one qubit can be tuned through the other with a strong suppression of state leakage. Finally, we investigate switching the tunable bus on nanosecond timescales. Pulsing the coupler has a similar signature to exciting the qubits albeit with suppressed phase coherence. The underlying mechanism behind this observation remains unclear.

%% Background
A schematic of the device is shown in Fig.~\ref{device}. Both qubits Q1 and Q2 are capacitively coupled to a $\lambda/2$ bus resonator with coupling strengths $g_{1(2)} \propto e\beta V_{\mathrm{rms,1(2)}}/\hbar$, where $\beta$ is the ratio of coupling capacitance to total qubit capacitance, and $V_{\mathrm{rms,1(2)}}$ is the root mean square of the zero-point voltage fluctuations of the resonator at the location of Q1(2)~\cite{wallraff_2004}. With the qubits at the same frequency $f_{\mathrm{Q1}},f_{\mathrm{Q2}}=f_{\mathrm{Q}}$, detuned by $\Delta=2\pi (f_{\mathrm{res}}-f_{\mathrm{Q}})$ from the resonator frequency $f_{\mathrm{res}}$, the bus-mediated qubit-qubit coupling $g_{12}={g_1 g_2}/\Delta$~\cite{blais_2004,sorensen_1999} can be controlled by changing either $\Delta$ or $g_{1}g_{2}$. 

An open switch gives a voltage antinode at the qubit end of the resonator [blue in Fig.~\ref{device}], which results in a large $V_{\mathrm{rms,1(2)}}$ with the resonator frequency given by the $\lambda/2$ mode, $f_{\mathrm{res}}=f_{\mathrm{\lambda/2}}$. With the SFET in this open state, and $f_{\mathrm{Q1}},f_{\mathrm{Q2}}$ close to $f_{\mathrm{\lambda/2}}$ we expect that the cavity-mediated coupling is turned on. On the other hand, when the switch is closed, a voltage node is enforced at the qubit end of the resonator, with its fundamental mode changing from $\lambda/2$ to $\lambda/4$. This turns off the interqubit coupling by reducing $V_{\mathrm{rms,1(2)}}$ and moving the lowest bus modes to $f_{\mathrm{\lambda/2}}/2$ and $3f_{\mathrm{\lambda/2}}/2$, which are far detuned from the qubit frequencies~\cite{wallquist_2006}.

\begin{figure}%aspect ratio: 258/128 ->  [(150 / ) + 20 words] = 95 words
    \centering
        \includegraphics[width=1\columnwidth]{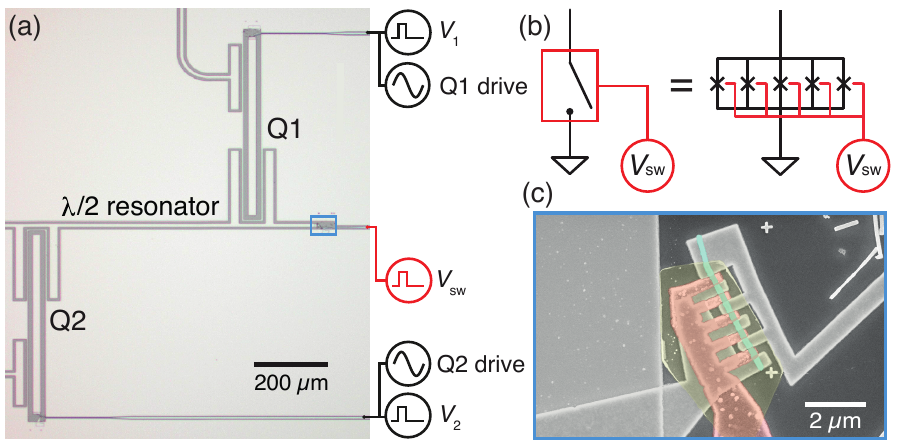}\vspace{-2mm}
    \caption{Two-qubit device with switchable quantum bus. (a) Optical micrograph of the two-gatemon device with the $\lambda$/2 bus resonator terminated by a superconducting switch. Each qubit consists of a bar-shaped island and a gated Al-InAs-Al Josephson junction. (b) The superconducting switch consists of five parallel gated semiconducting weak link Josephson junctions controlled by a single gate voltage. (c) Scanning electron micrograph of the five top gated Al-InAs-Al Josephson junctions.
    }
    \label{switch}
\end{figure}

\begin{figure}%aspect ratio: 243/260 ->  [(150 / ) + 20 words] = 181 words
    \centering
        \hspace{-2mm}\includegraphics[width=1\columnwidth]{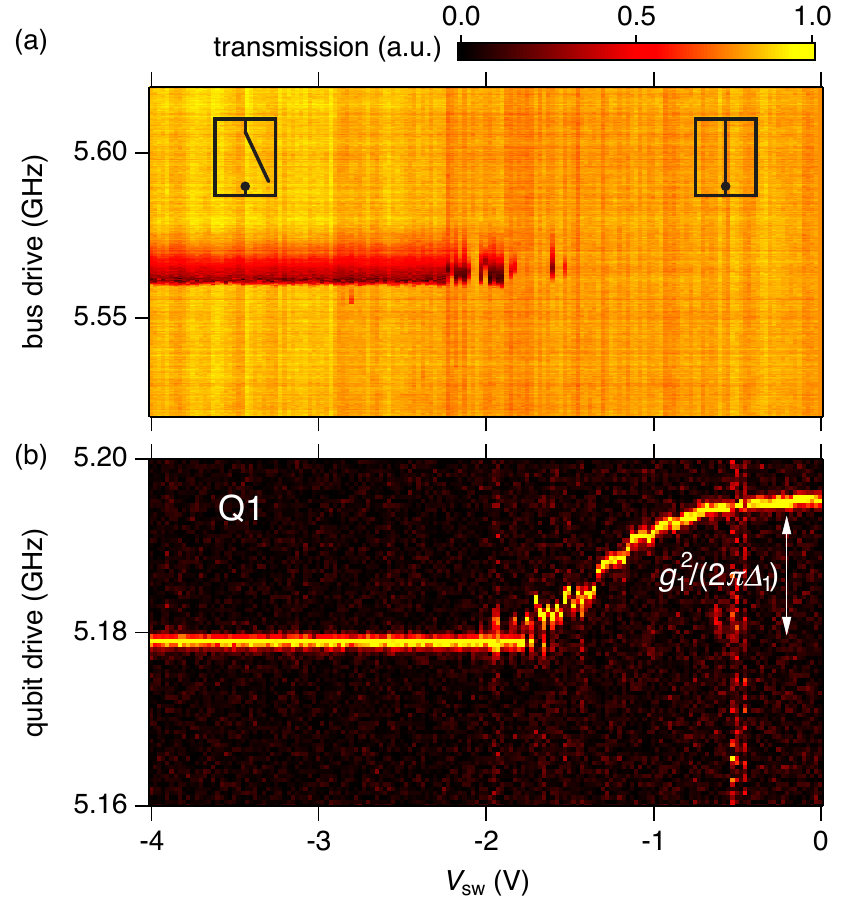}\vspace{-4mm}
    \caption{Switchable bus and qubit spectroscopy. (a) Normalized transmission as a function of bus drive frequency and $V_{\mathrm{sw}}$. (b) Q1 resonance frequency as a function of $V_{\mathrm{sw}}$. The Q1 readout resonator response was measured while a qubit microwave drive tone probed the Q1 transition frequency.
    }
    \label{spectroscopy}\vspace{-4mm}
\end{figure}

%% Methods
Figure~\ref{switch}(a) shows an optical image of the tunable bus device. The JJs for both the cavity and the qubits are superconductor-semiconductor-superconductor (S-Sm-S) junctions with a few-channel Sm region~\cite{kringhoj_2017}, allowing the Josephson coupling energy, $E_J$, to be tuned using a gate voltage that controls the carrier density in the Sm region. The two transmon-type gatemon qubits each consist of a bar-shaped island with a single JJ to ground. The SFET at the end of the tunable bus is made from several gate tunable JJs in parallel [Fig. \ref{switch}(b)].  

\begin{figure*} %aspect ratio: 569/294 -> {[300 / (0.5 * aspect ratio)] + 40 words} = 350 words 
\begin{center}
	\includegraphics[width=2\columnwidth]{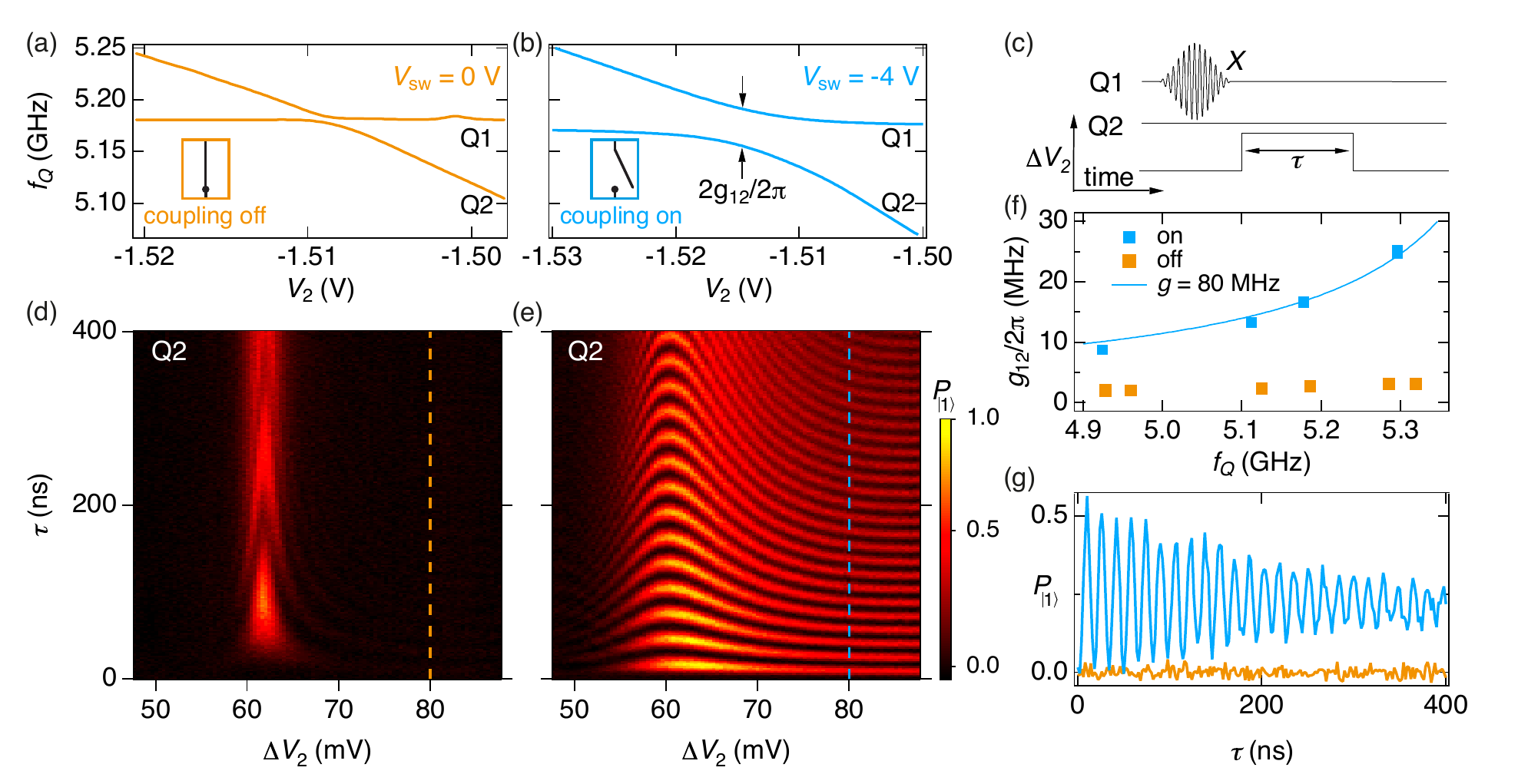}
\caption{\label{swap} Tunable coherent gatemon coupling. (a), [(b)] Measurement of the avoided level crossing between Q1 and Q2 for the switch closed (open), corresponding to gatemon coupling off (on) as a function of the qubit drive and $V_2$. (c) Pulse sequence to probe the coherent coupling between the qubits. With Q1 and Q2 detuned, Q1 is prepared in $|1\rangle$ and Q2 in $|0\rangle$. A square gate pulse with amplitude $\Delta V_2$ is turned on for a time $\tau$ and brings Q2 close to or in resonance with Q1. (d),[(e)] The $|1\rangle$ state probability, $P_{|1\rangle}$, for Q2 as a function of $\Delta V_2$ and $\tau$ for qubit coupling off (on). (f) Extracted gatemon coupling strengths for on and off case as a function of qubit resonance frequency. The solid line is a fit to the function $g_{\mathrm{12,on}}={g^2}/{\Delta}$. (f) Cuts along the dashed lines in (d) and (e) at $\Delta V_2=80$~mV.
}
\end{center}	
\vspace{-0.5cm}
\end{figure*}

The device was fabricated following the recipe described in Ref.~\cite{casparis_2016} and the supplementary material~\cite{sup}. Both the qubits and the tunable bus JJs were formed by selectively wet etching a segment of a $\sim$30~nm thick Al shell that was epitaxially grown around a $\sim$75~nm diameter single crystal InAs nanowire~\cite{krogstrup_2015}. $E_C/h$ of Q1(2) was designed to be $\sim$200 MHz with $E_J/E_C$ tuned to $75-90$ using the side gate voltage $V_{1(2)}$. To reduce the effective inductance of the bus switch when closed, five parallel JJs were used to form the SFET.  As shown in Fig.~\ref{switch}(b), the five junctions were etched into a single wire (blue) and then covered with 15~nm of ZrO\textsubscript{2} dielectric (yellow) deposited by atomic layer deposition. The SFET was controlled with a common top gate voltage $V_{\mathrm{sw}}$ (red).

The qubits were manipulated using phase-controlled microwave pulses for rotations around axes in the $X$-$Y$ plane of the Bloch sphere and voltage pulses on $V_{1,2}$ for rotations around the $Z$-axis and fast frequency displacement~\cite{rotations}. Both $X$-$Y$ and $Z$ control pulses were applied through each qubit's gate line. The two qubits were coupled to individual $\lambda$/4 superconducting cavities (with resonant frequencies $f_{\mathrm{C1}}\sim$~6.87~GHz and $f_{\mathrm{C2}}\sim$~6.80~GHz). These were then coupled to a common feedline for dispersive readout~\cite{barends_2013} with a superconducting travelling wave parametric amplifier used to improve the signal-to-noise ratio~\cite{macklin_2015}. The tunable bus was also coupled to the common feedline allowing an independent measurement of its resonance. The sample was placed inside an Al box, surrounded by a cryoperm shield and mounted at the mixing chamber of a cryogen-free dilution refrigerator with base temperature $\sim$~20~mK~\cite{sup}. 

% Results
Figure \ref{spectroscopy}(a) shows vector network analyzer measurements of the tunable bus resonance as a function of $V_{\mathrm{sw}}$. At large negative $V_{\mathrm{sw}}$, a resonance is observed at $f_{\mathrm{\lambda/2}}\sim$~5.6~GHz. Going to more positive $V_{\mathrm{sw}}$, the bus resonance disappears with some reentrant features indicating a non-monotonic turn on of the SFET. We speculate that the disappearance of the resonance is due to the measurement excitation populating the bus with photons and thus driving the SFET normal, leading to a highly reduced Q factor. Although affecting our ability to directly track the bus frequency, it should not impact its role as a quantum bus for Q1 and Q2 as the coupling is mediated through virtual photons~\cite{blais_2004}.
Interaction between the bus and the qubits renormalizes the qubit frequencies, allowing changes in the bus to be indirectly probed by measuring one of the qubits [Fig.~\ref{spectroscopy}(b)]. The push on $f_{\mathrm{Q1}}$ by the bus is given by the Lamb shift $\chi_{1}=g_{1}^2/(\Delta_{1})$~(white arrow), where $\Delta_1=2\pi (f_{\mathrm{res}}-f_{\mathrm{Q1}})$. When the SFET is depleted, the qubit frequency is pushed by the resonator with $f_{\mathrm{\lambda/2}}\sim$~5.6~GHz. While closing the switch $f_{\mathrm{Q1}}$ increased, indicating that either the bus mode is moving up in frequency or $g_1$ is decreased, or both. We observed a crossing of the readout resonator with the bus resonator at around $V_{\mathrm{sw}}=$~-0.5~V, characterized by a stripe in the spectroscopy data where the readout visibility is reduced. Both the continuous change of the qubit frequency and the crossing of a resonance with the readout resonator indicate that the first mode of the $\lambda$/2 resonator (switch open) turns continuously into the second mode of the $\lambda$/4 resonator (switch closed). For $V_{\mathrm{sw}}>-0.5$~V, the qubit frequency is roughly constant, indicating that either $f_{\mathrm{res}}$ no longer changes, or that $g_1$ is suppressed, although we cannot distinguish between these two effects.

Next, we turn to qubit coupling at fixed values of $V_{\mathrm{sw}}$ where the coupler is either on or off. We measured the spectrum while tuning Q2 into resonance with Q1 [Fig.~\ref{swap}(a) and (b)]. On resonance, the two-qubit states hybridize due to the bus-mediated coupling.  As Fig.~\ref{swap}(a) illustrates, the splitting was small, although clearly non-zero, when the switch is closed. For an open switch the qubit coupling significantly increased, resulting in a larger splitting between hybridized states [Fig.~\ref{swap}(b)].

To further investigate the interqubit coupling, we performed experiments in the time domain. The two qubits were detuned by $\sim$400~MHz and Q1 (Q2) was prepared in $|1\rangle$  ($|0\rangle$). A gate pulse was applied for a time $\tau$ to bring Q2 into resonance with Q1 [Fig.~\ref{swap}(c)]. Depending on $\tau$ and the pulse amplitude $\mathrm{\Delta}V_2$ elementary excitations swap between the two qubits. Figure~\ref{swap}(d) shows the swap oscillations with the coupler off and from sine fits to the oscillations, an interaction rate $g_{\mathrm{12}}^{\mathrm{off}}/2\pi\sim 3.2$~MHz is extracted, consistent with the avoided crossing measured in spectroscopy. With the coupler on, we observed significantly faster swap oscillations [Fig.~\ref{swap}(e)] and extract $g_{\mathrm{12}}^{\mathrm{on}}/2\pi\sim$18~MHz. 

Figure~\ref{swap}(f) plots the gatemon coupling strength extracted from swap oscillations as a function of qubit frequency. As expected, $g_{\mathrm{12}}^{\mathrm{on}}$ (blue) depended strongly on the detuning from the bus. Assuming $g_1=g_2=g$ and fitting the data to $g_{\mathrm{12}}^{\mathrm{on}}=g^2/\Delta$ yields $g/2\pi \sim$~80~MHz. While electrostatic simulations predict negligible direct capacitive coupling between the qubits, we measured a small residual off state coupling $g_{\mathrm{12}}^{\mathrm{off}}/2\pi\sim 2-4$~MHz, limiting the maximum on/off coupling ratio observed for this setup to $\sim 8$. The weak frequency dependence of $g_{\mathrm{12}}^{\mathrm{off}}$ suggests that it is dominated by coupling through spurious chip modes that may be suppressed through more careful microwave engineering, for example, by using airbridges~\cite{chen_air_2014}. Figure~\ref{swap}(g) shows cuts from Fig.~\ref{swap}(d)-(e) where the Q1 frequency crossed through the Q2 frequency and then back with the coupler either on or off. These data illustrate that even a modest switching ratio $g_{\mathrm{12}}^{\mathrm{on}}/g_{\mathrm{12}}^{\mathrm{off}}\sim 6$ allows both strong suppression of state leakage when the coupler is off and fast swaps when on. Comparing readout signals for the coupler in the on and off states, we extract an on/off leakage ratio of 65. This gives a lower bound for the leakage suppression as the off state signal is dominated by measurement noise. In the case of a Landau-Zener tunnelling process, a state leakage of $\sim$~50\% in the on state (blue) indicates a level velocity of $\sim$~20~MHz/ns. Since the level velocity is the same for both coupler states, one can estimate a state leakage of $\sim$~0.5\% in the coupler off state~\cite{LandauZener}.
 
\begin{figure}%aspect ratio: 258/171 ->  [(150 / ) + 20 words] = 120 words
    \centering
        \hspace{-2mm}\includegraphics[width=1\columnwidth]{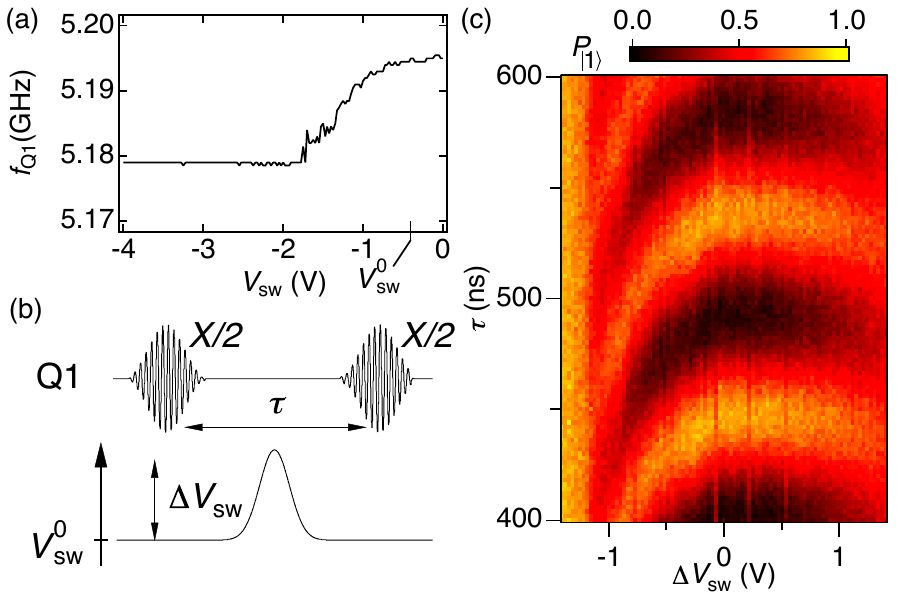}\vspace{-4mm}
    \caption{Fast switch response. (a) Resonance frequency of Q1 as a function of $V_{\mathrm{sw}}$, extracted from Fig.~\ref{spectroscopy}(b). (b) Ramsey pulse sequence to probe the fast response of the switch inserting a fast Gaussian switch pulse ($\sigma=64$ ns) with amplitude $\Delta V_{\mathrm{sw}}$ between two Ramsey pulses. (c) $P_{|1\rangle}$ as a function of $\Delta V_{\mathrm{sw}}$ and delay $\tau$. 
        }
    \label{fastpulse}\vspace{-4mm}
\end{figure}

Finally, we investigated dynamic operation of the switch by pulsing $V_{\mathrm{sw}}$. Figure~\ref{fastpulse}(a) shows the change of the qubit frequency $f_{\mathrm{Q1}}$ while controlling the bus. Again, $f_{\mathrm{Q1}}$ is pushed down at large negative $V_{\mathrm{sw}}$ due to the Lamb shift. We probed the effect that a fast voltage pulse on the switch has on Q1 through a Ramsey experiment. Two $X/2$ pulses were interleaved with a voltage pulse of the SFET gate [Fig.~\ref{fastpulse}(b)]. The Ramsey experiment is sensitive to the Lamb shift induced qubit frequency change. Sitting at a DC offset $V_{\mathrm{sw}}^0=$~-0.4~V, for $\Delta V_{\mathrm{sw}}>0$~V the Ramsey fringes remained roughly constant, as $f_{\mathrm{Q1}}$ does not change [Fig.~\ref{fastpulse}(c)]. At high pulse amplitudes the visibility of the fringes was reduced, indicating reduced qubit coherence. We speculate that above certain amplitudes charge traps in the gate dielectric are excited and only relax on time scales comparable to the Ramsey experiment, causing decoherence, though further experiments would be needed to verify this.

While applying negative pulses ($\Delta V_{\mathrm{sw}}<0$~V) to change the qubit coupling on a fast time scale, $f_{\mathrm{Q1}}$ was lowered, reducing the period of the Ramsey fringes. For the negative pulses above a certain critical amplitude, $\Delta V_{\mathrm{sw}}^{\mathrm{c}} \sim -1.1$~V, the readout response suggests that Q1 is excited into the $|1\rangle$ state and phase coherence is lost. The origin of this effect is presently unclear. We find that the value of $\Delta V_{\mathrm{sw}}^{\mathrm{c}}$ depends on both $V_{\mathrm{sw}}^0$ and the shape of the switch pulse. It was observed that the timescale on which the qubit can be coherently manipulated after a switch pulse is somewhat shorter than the decay time of the qubit, possibly indicating a different mechanism than qubit excitation like impairment of the readout resonator. Similar effects have been observed in two other samples, one device identical to that presented here, and the other using a $\lambda$/4 switchable resonator as the quantum bus. We speculate that pulsing the SFET close to depletion nonadiabatially excites the qubit circuit~\cite{shim_2016}. Another possibility is that pulsing the SFET JJs towards depletion generates quasiparticles that induce decoherence~\cite{patel_2016}. 

% Conclusions
In summary, we have demonstrated a voltage-tunable superconducting quantum bus that can control the coherent coupling between two gatemons. The number of qubit pairs coupled through the tunable resonator could readily be increased, allowing for larger connectivity. This could be of interest for qubit architectures beyond the surface code geometry~\cite{bartlett_2016}. Moreover, the continuously tunable coupling could prove attractive for quantum simulation~\cite{buluta_2009}. The potential advantages of tunable bus coupling motivate further investigation of how dynamic control of this voltage-controlled bus affects qubit operation.

% Acknowledgements
\begin{acknowledgments}
We acknowledge helpful discussions with S.~Nigg and V.~Shumeiko. This work was supported by Microsoft Project Q, the U.S. Army Research Office, and the Danish National Research Foundation. F.K acknowledges support from the Danish Innovation Fund  and C.M.M. acknowledges support from the Villum Foundation. The travelling wave parametric amplifier used in this experiment was provided by MIT Lincoln Laboratory and Irfan Siddiqi Quantum Consulting (ISQC), LLC via sponsorship from the US Government.
\end{acknowledgments}

%\bibliography{couplerbib}

\begin{thebibliography}{38}%
\makeatletter
\providecommand \@ifxundefined [1]{%
 \@ifx{#1\undefined}
}%
\providecommand \@ifnum [1]{%
 \ifnum #1\expandafter \@firstoftwo
 \else \expandafter \@secondoftwo
 \fi
}%
\providecommand \@ifx [1]{%
 \ifx #1\expandafter \@firstoftwo
 \else \expandafter \@secondoftwo
 \fi
}%
\providecommand \natexlab [1]{#1}%
\providecommand \enquote  [1]{``#1''}%
\providecommand \bibnamefont  [1]{#1}%
\providecommand \bibfnamefont [1]{#1}%
\providecommand \citenamefont [1]{#1}%
\providecommand \href@noop [0]{\@secondoftwo}%
\providecommand \href [0]{\begingroup \@sanitize@url \@href}%
\providecommand \@href[1]{\@@startlink{#1}\@@href}%
\providecommand \@@href[1]{\endgroup#1\@@endlink}%
\providecommand \@sanitize@url [0]{\catcode `\\12\catcode `\$12\catcode
  `\&12\catcode `\#12\catcode `\^12\catcode `\_12\catcode `\%12\relax}%
\providecommand \@@startlink[1]{}%
\providecommand \@@endlink[0]{}%
\providecommand \url  [0]{\begingroup\@sanitize@url \@url }%
\providecommand \@url [1]{\endgroup\@href {#1}{\urlprefix }}%
\providecommand \urlprefix  [0]{URL }%
\providecommand \Eprint [0]{\href }%
\providecommand \doibase [0]{http://dx.doi.org/}%
\providecommand \selectlanguage [0]{\@gobble}%
\providecommand \bibinfo  [0]{\@secondoftwo}%
\providecommand \bibfield  [0]{\@secondoftwo}%
\providecommand \translation [1]{[#1]}%
\providecommand \BibitemOpen [0]{}%
\providecommand \bibitemStop [0]{}%
\providecommand \bibitemNoStop [0]{.\EOS\space}%
\providecommand \EOS [0]{\spacefactor3000\relax}%
\providecommand \BibitemShut  [1]{\csname bibitem#1\endcsname}%
\let\auto@bib@innerbib\@empty
%</preamble>
\bibitem [{\citenamefont {DiCarlo}\ \emph {et~al.}(2009)\citenamefont
  {DiCarlo}, \citenamefont {Chow}, \citenamefont {Gambetta}, \citenamefont
  {Bishop}, \citenamefont {Johnson}, \citenamefont {Schuster}, \citenamefont
  {Majer}, \citenamefont {Blais}, \citenamefont {Frunzio}, \citenamefont
  {Girvin},\ and\ \citenamefont {Schoelkopf}}]{dicarlo_2009}%
  \BibitemOpen
  \bibfield  {author} {\bibinfo {author} {\bibfnamefont {L.}~\bibnamefont
  {DiCarlo}}, \bibinfo {author} {\bibfnamefont {J.~M.}\ \bibnamefont {Chow}},
  \bibinfo {author} {\bibfnamefont {J.~M.}\ \bibnamefont {Gambetta}}, \bibinfo
  {author} {\bibfnamefont {L.~S.}\ \bibnamefont {Bishop}}, \bibinfo {author}
  {\bibfnamefont {B.~R.}\ \bibnamefont {Johnson}}, \bibinfo {author}
  {\bibfnamefont {D.~I.}\ \bibnamefont {Schuster}}, \bibinfo {author}
  {\bibfnamefont {J.}~\bibnamefont {Majer}}, \bibinfo {author} {\bibfnamefont
  {A.}~\bibnamefont {Blais}}, \bibinfo {author} {\bibfnamefont
  {L.}~\bibnamefont {Frunzio}}, \bibinfo {author} {\bibfnamefont {S.~M.}\
  \bibnamefont {Girvin}}, \ and\ \bibinfo {author} {\bibfnamefont {R.~J.}\
  \bibnamefont {Schoelkopf}},\ }\href {\doibase 10.1038/nature08121} {\bibfield
   {journal} {\bibinfo  {journal} {Nature}\ }\textbf {\bibinfo {volume}
  {460}},\ \bibinfo {pages} {240} (\bibinfo {year} {2009})}\BibitemShut
  {NoStop}%
\bibitem [{\citenamefont {Kelly}\ \emph {et~al.}(2014)\citenamefont {Kelly},
  \citenamefont {Barends}, \citenamefont {Campbell}, \citenamefont {Chen},
  \citenamefont {Chen}, \citenamefont {Chiaro}, \citenamefont {Dunsworth},
  \citenamefont {Fowler}, \citenamefont {Hoi}, \citenamefont {Jeffrey},
  \citenamefont {Megrant}, \citenamefont {Mutus}, \citenamefont {Neill},
  \citenamefont {O'Malley}, \citenamefont {Quintana}, \citenamefont {Roushan},
  \citenamefont {Sank}, \citenamefont {Vainsencher}, \citenamefont {Wenner},
  \citenamefont {White}, \citenamefont {Cleland},\ and\ \citenamefont
  {Martinis}}]{kelly_2014}%
  \BibitemOpen
  \bibfield  {author} {\bibinfo {author} {\bibfnamefont {J.}~\bibnamefont
  {Kelly}}, \bibinfo {author} {\bibfnamefont {R.}~\bibnamefont {Barends}},
  \bibinfo {author} {\bibfnamefont {B.}~\bibnamefont {Campbell}}, \bibinfo
  {author} {\bibfnamefont {Y.}~\bibnamefont {Chen}}, \bibinfo {author}
  {\bibfnamefont {Z.}~\bibnamefont {Chen}}, \bibinfo {author} {\bibfnamefont
  {B.}~\bibnamefont {Chiaro}}, \bibinfo {author} {\bibfnamefont
  {A.}~\bibnamefont {Dunsworth}}, \bibinfo {author} {\bibfnamefont {A.~G.}\
  \bibnamefont {Fowler}}, \bibinfo {author} {\bibfnamefont {I.-C.}\
  \bibnamefont {Hoi}}, \bibinfo {author} {\bibfnamefont {E.}~\bibnamefont
  {Jeffrey}}, \bibinfo {author} {\bibfnamefont {A.}~\bibnamefont {Megrant}},
  \bibinfo {author} {\bibfnamefont {J.}~\bibnamefont {Mutus}}, \bibinfo
  {author} {\bibfnamefont {C.}~\bibnamefont {Neill}}, \bibinfo {author}
  {\bibfnamefont {P.~J.~J.}\ \bibnamefont {O'Malley}}, \bibinfo {author}
  {\bibfnamefont {C.}~\bibnamefont {Quintana}}, \bibinfo {author}
  {\bibfnamefont {P.}~\bibnamefont {Roushan}}, \bibinfo {author} {\bibfnamefont
  {D.}~\bibnamefont {Sank}}, \bibinfo {author} {\bibfnamefont {A.}~\bibnamefont
  {Vainsencher}}, \bibinfo {author} {\bibfnamefont {J.}~\bibnamefont {Wenner}},
  \bibinfo {author} {\bibfnamefont {T.~C.}\ \bibnamefont {White}}, \bibinfo
  {author} {\bibfnamefont {A.~N.}\ \bibnamefont {Cleland}}, \ and\ \bibinfo
  {author} {\bibfnamefont {J.~M.}\ \bibnamefont {Martinis}},\ }\href@noop {}
  {\bibfield  {journal} {\bibinfo  {journal} {Phys. Rev. Lett.}\ }\textbf
  {\bibinfo {volume} {112}},\ \bibinfo {pages} {240504} (\bibinfo {year}
  {2014})}\BibitemShut {NoStop}%
\bibitem [{\citenamefont {Majer}\ \emph {et~al.}(2007)\citenamefont {Majer},
  \citenamefont {Chow}, \citenamefont {Gambetta}, \citenamefont {Koch},
  \citenamefont {Johnson}, \citenamefont {Schreier}, \citenamefont {Frunzio},
  \citenamefont {Schuster}, \citenamefont {Houck}, \citenamefont {Wallraff},
  \citenamefont {Blais}, \citenamefont {Devoret}, \citenamefont {Girvin},\ and\
  \citenamefont {Schoelkopf}}]{majer_2007}%
  \BibitemOpen
  \bibfield  {author} {\bibinfo {author} {\bibfnamefont {J.}~\bibnamefont
  {Majer}}, \bibinfo {author} {\bibfnamefont {J.}~\bibnamefont {Chow}},
  \bibinfo {author} {\bibfnamefont {J.}~\bibnamefont {Gambetta}}, \bibinfo
  {author} {\bibfnamefont {J.}~\bibnamefont {Koch}}, \bibinfo {author}
  {\bibfnamefont {B.}~\bibnamefont {Johnson}}, \bibinfo {author} {\bibfnamefont
  {J.}~\bibnamefont {Schreier}}, \bibinfo {author} {\bibfnamefont
  {L.}~\bibnamefont {Frunzio}}, \bibinfo {author} {\bibfnamefont
  {D.}~\bibnamefont {Schuster}}, \bibinfo {author} {\bibfnamefont {A.~A.}\
  \bibnamefont {Houck}}, \bibinfo {author} {\bibfnamefont {A.}~\bibnamefont
  {Wallraff}}, \bibinfo {author} {\bibfnamefont {A.}~\bibnamefont {Blais}},
  \bibinfo {author} {\bibfnamefont {M.}~\bibnamefont {Devoret}}, \bibinfo
  {author} {\bibfnamefont {S.}~\bibnamefont {Girvin}}, \ and\ \bibinfo {author}
  {\bibfnamefont {R.}~\bibnamefont {Schoelkopf}},\ }\href@noop {} {\bibfield
  {journal} {\bibinfo  {journal} {Nature}\ }\textbf {\bibinfo {volume} {449}},\
  \bibinfo {pages} {443} (\bibinfo {year} {2007})}\BibitemShut {NoStop}%
\bibitem [{\citenamefont {Landau}(1932)}]{landau_1932}%
  \BibitemOpen
  \bibfield  {author} {\bibinfo {author} {\bibfnamefont {L.}~\bibnamefont
  {Landau}},\ }\href@noop {} {\bibfield  {journal} {\bibinfo  {journal} {Phys.
  Z. Sowjetunion}\ }\textbf {\bibinfo {volume} {2}},\ \bibinfo {pages} {46}
  (\bibinfo {year} {1932})}\BibitemShut {NoStop}%
\bibitem [{\citenamefont {Zener}(1932)}]{zener_1932}%
  \BibitemOpen
  \bibfield  {author} {\bibinfo {author} {\bibfnamefont {C.}~\bibnamefont
  {Zener}},\ }\href@noop {} {\bibfield  {journal} {\bibinfo  {journal} {Proc.
  R. Soc. A}\ }\textbf {\bibinfo {volume} {137}},\ \bibinfo {pages} {696}
  (\bibinfo {year} {1932})}\BibitemShut {NoStop}%
\bibitem [{\citenamefont {St{\"u}ckelberg}(1932)}]{stuckelberg_1932}%
  \BibitemOpen
  \bibfield  {author} {\bibinfo {author} {\bibfnamefont {E.~C.~G.}\
  \bibnamefont {St{\"u}ckelberg}},\ }\href@noop {} {\bibfield  {journal}
  {\bibinfo  {journal} {Helv. Phys. Acta}\ }\textbf {\bibinfo {volume} {5}},\
  \bibinfo {pages} {369} (\bibinfo {year} {1932})}\BibitemShut {NoStop}%
\bibitem [{\citenamefont {Majorana}(1932)}]{majorana_1932}%
  \BibitemOpen
  \bibfield  {author} {\bibinfo {author} {\bibfnamefont {E.}~\bibnamefont
  {Majorana}},\ }\href@noop {} {\bibfield  {journal} {\bibinfo  {journal}
  {Nuovo Cimento}\ }\textbf {\bibinfo {volume} {9}},\ \bibinfo {pages} {43}
  (\bibinfo {year} {1932})}\BibitemShut {NoStop}%
\bibitem [{\citenamefont {Quintana}\ \emph {et~al.}(2013)\citenamefont
  {Quintana}, \citenamefont {Petersson}, \citenamefont {McFaul}, \citenamefont
  {Srinivasan}, \citenamefont {Houck},\ and\ \citenamefont
  {Petta}}]{quintana_2013}%
  \BibitemOpen
  \bibfield  {author} {\bibinfo {author} {\bibfnamefont {C.~M.}\ \bibnamefont
  {Quintana}}, \bibinfo {author} {\bibfnamefont {K.~D.}\ \bibnamefont
  {Petersson}}, \bibinfo {author} {\bibfnamefont {L.~W.}\ \bibnamefont
  {McFaul}}, \bibinfo {author} {\bibfnamefont {S.~J.}\ \bibnamefont
  {Srinivasan}}, \bibinfo {author} {\bibfnamefont {A.~A.}\ \bibnamefont
  {Houck}}, \ and\ \bibinfo {author} {\bibfnamefont {J.~R.}\ \bibnamefont
  {Petta}},\ }\href@noop {} {\bibfield  {journal} {\bibinfo  {journal} {Phys.
  Rev. Lett.}\ }\textbf {\bibinfo {volume} {110}},\ \bibinfo {pages} {173603}
  (\bibinfo {year} {2013})}\BibitemShut {NoStop}%
\bibitem [{\citenamefont {Bialczak}\ \emph {et~al.}(2011)\citenamefont
  {Bialczak}, \citenamefont {Ansmann}, \citenamefont {Hofheinz}, \citenamefont
  {Lenander}, \citenamefont {Lucero}, \citenamefont {Neeley}, \citenamefont
  {O'Connell}, \citenamefont {Sank}, \citenamefont {Wang}, \citenamefont
  {Weides}, \citenamefont {Wenner}, \citenamefont {Yamamoto}, \citenamefont
  {Cleland},\ and\ \citenamefont {Martinis}}]{bialczak_2011}%
  \BibitemOpen
  \bibfield  {author} {\bibinfo {author} {\bibfnamefont {R.~C.}\ \bibnamefont
  {Bialczak}}, \bibinfo {author} {\bibfnamefont {M.}~\bibnamefont {Ansmann}},
  \bibinfo {author} {\bibfnamefont {M.}~\bibnamefont {Hofheinz}}, \bibinfo
  {author} {\bibfnamefont {M.}~\bibnamefont {Lenander}}, \bibinfo {author}
  {\bibfnamefont {E.}~\bibnamefont {Lucero}}, \bibinfo {author} {\bibfnamefont
  {M.}~\bibnamefont {Neeley}}, \bibinfo {author} {\bibfnamefont {A.~D.}\
  \bibnamefont {O'Connell}}, \bibinfo {author} {\bibfnamefont {D.}~\bibnamefont
  {Sank}}, \bibinfo {author} {\bibfnamefont {H.}~\bibnamefont {Wang}}, \bibinfo
  {author} {\bibfnamefont {M.}~\bibnamefont {Weides}}, \bibinfo {author}
  {\bibfnamefont {J.}~\bibnamefont {Wenner}}, \bibinfo {author} {\bibfnamefont
  {T.}~\bibnamefont {Yamamoto}}, \bibinfo {author} {\bibfnamefont {A.~N.}\
  \bibnamefont {Cleland}}, \ and\ \bibinfo {author} {\bibfnamefont {J.~M.}\
  \bibnamefont {Martinis}},\ }\href@noop {} {\bibfield  {journal} {\bibinfo
  {journal} {Phys. Rev. Lett.}\ }\textbf {\bibinfo {volume} {106}},\ \bibinfo
  {pages} {060501} (\bibinfo {year} {2011})}\BibitemShut {NoStop}%
\bibitem [{\citenamefont {Chen}\ \emph
  {et~al.}(2014{\natexlab{a}})\citenamefont {Chen}, \citenamefont {Neill},
  \citenamefont {Roushan}, \citenamefont {Leung}, \citenamefont {Fang},
  \citenamefont {Barends}, \citenamefont {Kelly}, \citenamefont {Campbell},
  \citenamefont {Chen}, \citenamefont {Chiaro}, \citenamefont {Dunsworth},
  \citenamefont {Jeffrey}, \citenamefont {Megrant}, \citenamefont {Mutus},
  \citenamefont {O'Malley}, \citenamefont {Quintana}, \citenamefont {Sank},
  \citenamefont {Vainsencher}, \citenamefont {Wenner}, \citenamefont {White},
  \citenamefont {Geller}, \citenamefont {Cleland},\ and\ \citenamefont
  {Martinis}}]{chen_2014}%
  \BibitemOpen
  \bibfield  {author} {\bibinfo {author} {\bibfnamefont {Y.}~\bibnamefont
  {Chen}}, \bibinfo {author} {\bibfnamefont {C.}~\bibnamefont {Neill}},
  \bibinfo {author} {\bibfnamefont {P.}~\bibnamefont {Roushan}}, \bibinfo
  {author} {\bibfnamefont {N.}~\bibnamefont {Leung}}, \bibinfo {author}
  {\bibfnamefont {M.}~\bibnamefont {Fang}}, \bibinfo {author} {\bibfnamefont
  {R.}~\bibnamefont {Barends}}, \bibinfo {author} {\bibfnamefont
  {J.}~\bibnamefont {Kelly}}, \bibinfo {author} {\bibfnamefont
  {B.}~\bibnamefont {Campbell}}, \bibinfo {author} {\bibfnamefont
  {Z.}~\bibnamefont {Chen}}, \bibinfo {author} {\bibfnamefont {B.}~\bibnamefont
  {Chiaro}}, \bibinfo {author} {\bibfnamefont {A.}~\bibnamefont {Dunsworth}},
  \bibinfo {author} {\bibfnamefont {E.}~\bibnamefont {Jeffrey}}, \bibinfo
  {author} {\bibfnamefont {A.}~\bibnamefont {Megrant}}, \bibinfo {author}
  {\bibfnamefont {J.~Y.}\ \bibnamefont {Mutus}}, \bibinfo {author}
  {\bibfnamefont {P.~J.~J.}\ \bibnamefont {O'Malley}}, \bibinfo {author}
  {\bibfnamefont {C.~M.}\ \bibnamefont {Quintana}}, \bibinfo {author}
  {\bibfnamefont {D.}~\bibnamefont {Sank}}, \bibinfo {author} {\bibfnamefont
  {A.}~\bibnamefont {Vainsencher}}, \bibinfo {author} {\bibfnamefont
  {J.}~\bibnamefont {Wenner}}, \bibinfo {author} {\bibfnamefont {T.~C.}\
  \bibnamefont {White}}, \bibinfo {author} {\bibfnamefont {M.~R.}\ \bibnamefont
  {Geller}}, \bibinfo {author} {\bibfnamefont {A.~N.}\ \bibnamefont {Cleland}},
  \ and\ \bibinfo {author} {\bibfnamefont {J.~M.}\ \bibnamefont {Martinis}},\
  }\href@noop {} {\bibfield  {journal} {\bibinfo  {journal} {Phys. Rev. Lett.}\
  }\textbf {\bibinfo {volume} {113}},\ \bibinfo {pages} {220502} (\bibinfo
  {year} {2014}{\natexlab{a}})}\BibitemShut {NoStop}%
\bibitem [{\citenamefont {McKay}\ \emph {et~al.}(2016)\citenamefont {McKay},
  \citenamefont {Filipp}, \citenamefont {Mezzacapo}, \citenamefont {Magesan},
  \citenamefont {Chow},\ and\ \citenamefont {Gambetta}}]{mckay_2016}%
  \BibitemOpen
  \bibfield  {author} {\bibinfo {author} {\bibfnamefont {D.~C.}\ \bibnamefont
  {McKay}}, \bibinfo {author} {\bibfnamefont {S.}~\bibnamefont {Filipp}},
  \bibinfo {author} {\bibfnamefont {A.}~\bibnamefont {Mezzacapo}}, \bibinfo
  {author} {\bibfnamefont {E.}~\bibnamefont {Magesan}}, \bibinfo {author}
  {\bibfnamefont {J.~M.}\ \bibnamefont {Chow}}, \ and\ \bibinfo {author}
  {\bibfnamefont {J.~M.}\ \bibnamefont {Gambetta}},\ }\href {\doibase
  10.1103/PhysRevApplied.6.064007} {\bibfield  {journal} {\bibinfo  {journal}
  {Phys. Rev. Applied}\ }\textbf {\bibinfo {volume} {6}},\ \bibinfo {pages}
  {064007} (\bibinfo {year} {2016})}\BibitemShut {NoStop}%
\bibitem [{\citenamefont {Wallquist}\ \emph {et~al.}(2006)\citenamefont
  {Wallquist}, \citenamefont {Shumeiko},\ and\ \citenamefont
  {Wendin}}]{wallquist_2006}%
  \BibitemOpen
  \bibfield  {author} {\bibinfo {author} {\bibfnamefont {M.}~\bibnamefont
  {Wallquist}}, \bibinfo {author} {\bibfnamefont {V.~S.}\ \bibnamefont
  {Shumeiko}}, \ and\ \bibinfo {author} {\bibfnamefont {G.}~\bibnamefont
  {Wendin}},\ }\href@noop {} {\bibfield  {journal} {\bibinfo  {journal} {Phys.
  Rev. B}\ }\textbf {\bibinfo {volume} {74}},\ \bibinfo {pages} {224506}
  (\bibinfo {year} {2006})}\BibitemShut {NoStop}%
\bibitem [{\citenamefont {Wallraff}\ \emph {et~al.}(2004)\citenamefont
  {Wallraff}, \citenamefont {Schuster}, \citenamefont {Blais}, \citenamefont
  {Frunzio}, \citenamefont {Huang}, \citenamefont {Majer}, \citenamefont
  {Kumar}, \citenamefont {Girvin},\ and\ \citenamefont
  {Schoelkopf}}]{wallraff_2004}%
  \BibitemOpen
  \bibfield  {author} {\bibinfo {author} {\bibfnamefont {A.}~\bibnamefont
  {Wallraff}}, \bibinfo {author} {\bibfnamefont {D.~I.}\ \bibnamefont
  {Schuster}}, \bibinfo {author} {\bibfnamefont {A.}~\bibnamefont {Blais}},
  \bibinfo {author} {\bibfnamefont {L.}~\bibnamefont {Frunzio}}, \bibinfo
  {author} {\bibfnamefont {R.-S.}\ \bibnamefont {Huang}}, \bibinfo {author}
  {\bibfnamefont {J.}~\bibnamefont {Majer}}, \bibinfo {author} {\bibfnamefont
  {S.}~\bibnamefont {Kumar}}, \bibinfo {author} {\bibfnamefont {S.~M.}\
  \bibnamefont {Girvin}}, \ and\ \bibinfo {author} {\bibfnamefont {R.~J.}\
  \bibnamefont {Schoelkopf}},\ }\href {\doibase 10.1038/nature02851} {\bibfield
   {journal} {\bibinfo  {journal} {Nature}\ }\textbf {\bibinfo {volume}
  {431}},\ \bibinfo {pages} {162} (\bibinfo {year} {2004})}\BibitemShut
  {NoStop}%
\bibitem [{\citenamefont {Sillanp{\"a}{\"a}}\ \emph {et~al.}(2007)\citenamefont
  {Sillanp{\"a}{\"a}}, \citenamefont {Park},\ and\ \citenamefont
  {Simmonds}}]{sillanpaa_2007}%
  \BibitemOpen
  \bibfield  {author} {\bibinfo {author} {\bibfnamefont {M.~A.}\ \bibnamefont
  {Sillanp{\"a}{\"a}}}, \bibinfo {author} {\bibfnamefont {J.~I.}\ \bibnamefont
  {Park}}, \ and\ \bibinfo {author} {\bibfnamefont {R.~W.}\ \bibnamefont
  {Simmonds}},\ }\href@noop {} {\bibfield  {journal} {\bibinfo  {journal}
  {Nature}\ }\textbf {\bibinfo {volume} {449}},\ \bibinfo {pages} {438}
  (\bibinfo {year} {2007})}\BibitemShut {NoStop}%
\bibitem [{\citenamefont {Sandberg}\ \emph {et~al.}(2008)\citenamefont
  {Sandberg}, \citenamefont {Wilson}, \citenamefont {Persson}, \citenamefont
  {Bauch}, \citenamefont {Johansson}, \citenamefont {Shumeiko}, \citenamefont
  {Duty},\ and\ \citenamefont {Delsing}}]{sandberg_2008}%
  \BibitemOpen
  \bibfield  {author} {\bibinfo {author} {\bibfnamefont {M.}~\bibnamefont
  {Sandberg}}, \bibinfo {author} {\bibfnamefont {C.~M.}\ \bibnamefont
  {Wilson}}, \bibinfo {author} {\bibfnamefont {F.}~\bibnamefont {Persson}},
  \bibinfo {author} {\bibfnamefont {T.}~\bibnamefont {Bauch}}, \bibinfo
  {author} {\bibfnamefont {G.}~\bibnamefont {Johansson}}, \bibinfo {author}
  {\bibfnamefont {V.}~\bibnamefont {Shumeiko}}, \bibinfo {author}
  {\bibfnamefont {T.}~\bibnamefont {Duty}}, \ and\ \bibinfo {author}
  {\bibfnamefont {P.}~\bibnamefont {Delsing}},\ }\href {\doibase
  10.1063/1.2929367} {\bibfield  {journal} {\bibinfo  {journal} {Appl. Phys.
  Lett.}\ }\textbf {\bibinfo {volume} {92}},\ \bibinfo {pages} {203501}
  (\bibinfo {year} {2008})}\BibitemShut {NoStop}%
\bibitem [{\citenamefont {Palacios-Laloy}\ \emph {et~al.}(2008)\citenamefont
  {Palacios-Laloy}, \citenamefont {Nguyen}, \citenamefont {Mallet},
  \citenamefont {Bertet}, \citenamefont {Vion},\ and\ \citenamefont
  {Esteve}}]{palacios_2008}%
  \BibitemOpen
  \bibfield  {author} {\bibinfo {author} {\bibfnamefont {A.}~\bibnamefont
  {Palacios-Laloy}}, \bibinfo {author} {\bibfnamefont {F.}~\bibnamefont
  {Nguyen}}, \bibinfo {author} {\bibfnamefont {F.}~\bibnamefont {Mallet}},
  \bibinfo {author} {\bibfnamefont {P.}~\bibnamefont {Bertet}}, \bibinfo
  {author} {\bibfnamefont {D.}~\bibnamefont {Vion}}, \ and\ \bibinfo {author}
  {\bibfnamefont {D.}~\bibnamefont {Esteve}},\ }\href {\doibase
  10.1007/s10909-008-9774-x} {\bibfield  {journal} {\bibinfo  {journal} {J. Low
  Temp. Phys.}\ }\textbf {\bibinfo {volume} {151}},\ \bibinfo {pages} {1034}
  (\bibinfo {year} {2008})}\BibitemShut {NoStop}%
\bibitem [{\citenamefont {Kubo}\ \emph {et~al.}(2011)\citenamefont {Kubo},
  \citenamefont {Grezes}, \citenamefont {Dewes}, \citenamefont {Umeda},
  \citenamefont {Isoya}, \citenamefont {Sumiya}, \citenamefont {Morishita},
  \citenamefont {Abe}, \citenamefont {Onoda}, \citenamefont {Ohshima},
  \citenamefont {Jacques}, \citenamefont {Dr{\'e}au}, \citenamefont {Roch},
  \citenamefont {Diniz}, \citenamefont {Auffeves}, \citenamefont {Vion},
  \citenamefont {Esteve},\ and\ \citenamefont {Bertet}}]{kubo_2011}%
  \BibitemOpen
  \bibfield  {author} {\bibinfo {author} {\bibfnamefont {Y.}~\bibnamefont
  {Kubo}}, \bibinfo {author} {\bibfnamefont {C.}~\bibnamefont {Grezes}},
  \bibinfo {author} {\bibfnamefont {A.}~\bibnamefont {Dewes}}, \bibinfo
  {author} {\bibfnamefont {T.}~\bibnamefont {Umeda}}, \bibinfo {author}
  {\bibfnamefont {J.}~\bibnamefont {Isoya}}, \bibinfo {author} {\bibfnamefont
  {H.}~\bibnamefont {Sumiya}}, \bibinfo {author} {\bibfnamefont
  {N.}~\bibnamefont {Morishita}}, \bibinfo {author} {\bibfnamefont
  {H.}~\bibnamefont {Abe}}, \bibinfo {author} {\bibfnamefont {S.}~\bibnamefont
  {Onoda}}, \bibinfo {author} {\bibfnamefont {T.}~\bibnamefont {Ohshima}},
  \bibinfo {author} {\bibfnamefont {V.}~\bibnamefont {Jacques}}, \bibinfo
  {author} {\bibfnamefont {A.}~\bibnamefont {Dr{\'e}au}}, \bibinfo {author}
  {\bibfnamefont {J.-F.}\ \bibnamefont {Roch}}, \bibinfo {author}
  {\bibfnamefont {I.}~\bibnamefont {Diniz}}, \bibinfo {author} {\bibfnamefont
  {A.}~\bibnamefont {Auffeves}}, \bibinfo {author} {\bibfnamefont
  {D.}~\bibnamefont {Vion}}, \bibinfo {author} {\bibfnamefont {D.}~\bibnamefont
  {Esteve}}, \ and\ \bibinfo {author} {\bibfnamefont {P.}~\bibnamefont
  {Bertet}},\ }\href@noop {} {\bibfield  {journal} {\bibinfo  {journal} {Phys.
  Rev. Lett.}\ }\textbf {\bibinfo {volume} {107}},\ \bibinfo {pages} {220501}
  (\bibinfo {year} {2011})}\BibitemShut {NoStop}%
\bibitem [{\citenamefont {de~Lange}\ \emph {et~al.}(2015)\citenamefont
  {de~Lange}, \citenamefont {van Heck}, \citenamefont {Bruno}, \citenamefont
  {van Woerkom}, \citenamefont {Geresdi}, \citenamefont {Plissard},
  \citenamefont {Bakkers}, \citenamefont {Akhmerov},\ and\ \citenamefont
  {DiCarlo}}]{delange_2015}%
  \BibitemOpen
  \bibfield  {author} {\bibinfo {author} {\bibfnamefont {G.}~\bibnamefont
  {de~Lange}}, \bibinfo {author} {\bibfnamefont {B.}~\bibnamefont {van Heck}},
  \bibinfo {author} {\bibfnamefont {A.}~\bibnamefont {Bruno}}, \bibinfo
  {author} {\bibfnamefont {D.~J.}\ \bibnamefont {van Woerkom}}, \bibinfo
  {author} {\bibfnamefont {A.}~\bibnamefont {Geresdi}}, \bibinfo {author}
  {\bibfnamefont {S.~R.}\ \bibnamefont {Plissard}}, \bibinfo {author}
  {\bibfnamefont {E.~P. A.~M.}\ \bibnamefont {Bakkers}}, \bibinfo {author}
  {\bibfnamefont {A.~R.}\ \bibnamefont {Akhmerov}}, \ and\ \bibinfo {author}
  {\bibfnamefont {L.}~\bibnamefont {DiCarlo}},\ }\href@noop {} {\bibfield
  {journal} {\bibinfo  {journal} {Phys. Rev. Lett.}\ }\textbf {\bibinfo
  {volume} {115}},\ \bibinfo {pages} {127002} (\bibinfo {year}
  {2015})}\BibitemShut {NoStop}%
\bibitem [{\citenamefont {Larsen}\ \emph {et~al.}(2015)\citenamefont {Larsen},
  \citenamefont {Petersson}, \citenamefont {Kuemmeth}, \citenamefont
  {Jespersen}, \citenamefont {Krogstrup}, \citenamefont {Nygard},\ and\
  \citenamefont {Marcus}}]{larsen_2015}%
  \BibitemOpen
  \bibfield  {author} {\bibinfo {author} {\bibfnamefont {T.~W.}\ \bibnamefont
  {Larsen}}, \bibinfo {author} {\bibfnamefont {K.~D.}\ \bibnamefont
  {Petersson}}, \bibinfo {author} {\bibfnamefont {F.}~\bibnamefont {Kuemmeth}},
  \bibinfo {author} {\bibfnamefont {T.~S.}\ \bibnamefont {Jespersen}}, \bibinfo
  {author} {\bibfnamefont {P.}~\bibnamefont {Krogstrup}}, \bibinfo {author}
  {\bibfnamefont {J.}~\bibnamefont {Nygard}}, \ and\ \bibinfo {author}
  {\bibfnamefont {C.~M.}\ \bibnamefont {Marcus}},\ }\href@noop {} {\bibfield
  {journal} {\bibinfo  {journal} {Phys. Rev. Lett.}\ }\textbf {\bibinfo
  {volume} {115}},\ \bibinfo {pages} {127001} (\bibinfo {year}
  {2015})}\BibitemShut {NoStop}%
\bibitem [{\citenamefont {Ward}\ \emph {et~al.}(2013)\citenamefont {Ward},
  \citenamefont {Savage}, \citenamefont {Lagally}, \citenamefont
  {Coppersmith},\ and\ \citenamefont {Eriksson}}]{ward_2013}%
  \BibitemOpen
  \bibfield  {author} {\bibinfo {author} {\bibfnamefont {D.~R.}\ \bibnamefont
  {Ward}}, \bibinfo {author} {\bibfnamefont {D.~E.}\ \bibnamefont {Savage}},
  \bibinfo {author} {\bibfnamefont {M.~G.}\ \bibnamefont {Lagally}}, \bibinfo
  {author} {\bibfnamefont {S.~N.}\ \bibnamefont {Coppersmith}}, \ and\ \bibinfo
  {author} {\bibfnamefont {M.~A.}\ \bibnamefont {Eriksson}},\ }\href {\doibase
  10.1063/1.4807768} {\bibfield  {journal} {\bibinfo  {journal} {Appl. Phys.
  Lett.}\ }\textbf {\bibinfo {volume} {102}},\ \bibinfo {pages} {213107}
  (\bibinfo {year} {2013})}\BibitemShut {NoStop}%
\bibitem [{\citenamefont {Al-Taie}\ \emph {et~al.}(2013)\citenamefont
  {Al-Taie}, \citenamefont {Smith}, \citenamefont {Xu}, \citenamefont {See},
  \citenamefont {Griffiths}, \citenamefont {Beere}, \citenamefont {Jones},
  \citenamefont {Ritchie}, \citenamefont {Kelly},\ and\ \citenamefont
  {Smith}}]{al-taie_2013}%
  \BibitemOpen
  \bibfield  {author} {\bibinfo {author} {\bibfnamefont {H.}~\bibnamefont
  {Al-Taie}}, \bibinfo {author} {\bibfnamefont {L.~W.}\ \bibnamefont {Smith}},
  \bibinfo {author} {\bibfnamefont {B.}~\bibnamefont {Xu}}, \bibinfo {author}
  {\bibfnamefont {P.}~\bibnamefont {See}}, \bibinfo {author} {\bibfnamefont
  {J.~P.}\ \bibnamefont {Griffiths}}, \bibinfo {author} {\bibfnamefont {H.~E.}\
  \bibnamefont {Beere}}, \bibinfo {author} {\bibfnamefont {G.~A.~C.}\
  \bibnamefont {Jones}}, \bibinfo {author} {\bibfnamefont {D.~A.}\ \bibnamefont
  {Ritchie}}, \bibinfo {author} {\bibfnamefont {M.~J.}\ \bibnamefont {Kelly}},
  \ and\ \bibinfo {author} {\bibfnamefont {C.~G.}\ \bibnamefont {Smith}},\
  }\href {\doibase 10.1063/1.4811376} {\bibfield  {journal} {\bibinfo
  {journal} {Appl. Phys. Lett.}\ }\textbf {\bibinfo {volume} {102}},\ \bibinfo
  {pages} {243102} (\bibinfo {year} {2013})}\BibitemShut {NoStop}%
\bibitem [{\citenamefont {Hornibrook}\ \emph {et~al.}(2015)\citenamefont
  {Hornibrook}, \citenamefont {Colless}, \citenamefont {Conway~Lamb},
  \citenamefont {Pauka}, \citenamefont {Lu}, \citenamefont {Gossard},
  \citenamefont {Watson}, \citenamefont {Gardner}, \citenamefont {Fallahi},
  \citenamefont {Manfra},\ and\ \citenamefont {Reilly}}]{hornibrook_2014}%
  \BibitemOpen
  \bibfield  {author} {\bibinfo {author} {\bibfnamefont {J.~M.}\ \bibnamefont
  {Hornibrook}}, \bibinfo {author} {\bibfnamefont {J.~I.}\ \bibnamefont
  {Colless}}, \bibinfo {author} {\bibfnamefont {I.~D.}\ \bibnamefont
  {Conway~Lamb}}, \bibinfo {author} {\bibfnamefont {S.~J.}\ \bibnamefont
  {Pauka}}, \bibinfo {author} {\bibfnamefont {H.}~\bibnamefont {Lu}}, \bibinfo
  {author} {\bibfnamefont {A.~C.}\ \bibnamefont {Gossard}}, \bibinfo {author}
  {\bibfnamefont {J.~D.}\ \bibnamefont {Watson}}, \bibinfo {author}
  {\bibfnamefont {G.~C.}\ \bibnamefont {Gardner}}, \bibinfo {author}
  {\bibfnamefont {S.}~\bibnamefont {Fallahi}}, \bibinfo {author} {\bibfnamefont
  {M.~J.}\ \bibnamefont {Manfra}}, \ and\ \bibinfo {author} {\bibfnamefont
  {D.~J.}\ \bibnamefont {Reilly}},\ }\href@noop {} {\bibfield  {journal}
  {\bibinfo  {journal} {Phys. Rev. Applied}\ }\textbf {\bibinfo {volume} {3}},\
  \bibinfo {pages} {024010} (\bibinfo {year} {2015})}\BibitemShut {NoStop}%
\bibitem [{\citenamefont {Qi}\ \emph {et~al.}(2018)\citenamefont {Qi},
  \citenamefont {Xie}, \citenamefont {Shabani}, \citenamefont {Manucharyan},
  \citenamefont {Levchenko},\ and\ \citenamefont {Vavilov}}]{Qi_2018}%
  \BibitemOpen
  \bibfield  {author} {\bibinfo {author} {\bibfnamefont {Z.}~\bibnamefont
  {Qi}}, \bibinfo {author} {\bibfnamefont {H.-Y.}\ \bibnamefont {Xie}},
  \bibinfo {author} {\bibfnamefont {J.}~\bibnamefont {Shabani}}, \bibinfo
  {author} {\bibfnamefont {V.~E.}\ \bibnamefont {Manucharyan}}, \bibinfo
  {author} {\bibfnamefont {A.}~\bibnamefont {Levchenko}}, \ and\ \bibinfo
  {author} {\bibfnamefont {M.~G.}\ \bibnamefont {Vavilov}},\ }\href
  {http://arxiv.org/abs/1801.04291} {\bibfield  {journal} {\bibinfo  {journal}
  {arXiv preprint arXiv:1801.04291}\ } (\bibinfo {year} {2018})}\BibitemShut
  {NoStop}%
\bibitem [{\citenamefont {Blais}\ \emph {et~al.}(2004)\citenamefont {Blais},
  \citenamefont {Huang}, \citenamefont {Wallraff}, \citenamefont {Girvin},\
  and\ \citenamefont {Schoelkopf}}]{blais_2004}%
  \BibitemOpen
  \bibfield  {author} {\bibinfo {author} {\bibfnamefont {A.}~\bibnamefont
  {Blais}}, \bibinfo {author} {\bibfnamefont {R.-S.}\ \bibnamefont {Huang}},
  \bibinfo {author} {\bibfnamefont {A.}~\bibnamefont {Wallraff}}, \bibinfo
  {author} {\bibfnamefont {S.~M.}\ \bibnamefont {Girvin}}, \ and\ \bibinfo
  {author} {\bibfnamefont {R.~J.}\ \bibnamefont {Schoelkopf}},\ }\href@noop {}
  {\bibfield  {journal} {\bibinfo  {journal} {Phys. Rev. A}\ }\textbf {\bibinfo
  {volume} {69}},\ \bibinfo {pages} {062320} (\bibinfo {year}
  {2004})}\BibitemShut {NoStop}%
\bibitem [{\citenamefont {S{\o}rensen}\ and\ \citenamefont
  {M{\o}lmer}(1999)}]{sorensen_1999}%
  \BibitemOpen
  \bibfield  {author} {\bibinfo {author} {\bibfnamefont {A.}~\bibnamefont
  {S{\o}rensen}}\ and\ \bibinfo {author} {\bibfnamefont {K.}~\bibnamefont
  {M{\o}lmer}},\ }\href@noop {} {\bibfield  {journal} {\bibinfo  {journal}
  {Phys. Rev. Lett.}\ }\textbf {\bibinfo {volume} {82}},\ \bibinfo {pages}
  {1971} (\bibinfo {year} {1999})}\BibitemShut {NoStop}%
\bibitem [{\citenamefont {Kringh{\o}j}\ \emph {et~al.}(2017)\citenamefont
  {Kringh{\o}j}, \citenamefont {Casparis}, \citenamefont {Hell}, \citenamefont
  {Larsen}, \citenamefont {Kuemmeth}, \citenamefont {Leijnse}, \citenamefont
  {Flensberg}, \citenamefont {Krogstrup}, \citenamefont {Nyg{\aa}rd},
  \citenamefont {Petersson},\ and\ \citenamefont {Marcus}}]{kringhoj_2017}%
  \BibitemOpen
  \bibfield  {author} {\bibinfo {author} {\bibfnamefont {A.}~\bibnamefont
  {Kringh{\o}j}}, \bibinfo {author} {\bibfnamefont {L.}~\bibnamefont
  {Casparis}}, \bibinfo {author} {\bibfnamefont {M.}~\bibnamefont {Hell}},
  \bibinfo {author} {\bibfnamefont {T.~W.}\ \bibnamefont {Larsen}}, \bibinfo
  {author} {\bibfnamefont {F.}~\bibnamefont {Kuemmeth}}, \bibinfo {author}
  {\bibfnamefont {M.}~\bibnamefont {Leijnse}}, \bibinfo {author} {\bibfnamefont
  {K.}~\bibnamefont {Flensberg}}, \bibinfo {author} {\bibfnamefont
  {P.}~\bibnamefont {Krogstrup}}, \bibinfo {author} {\bibfnamefont
  {J.}~\bibnamefont {Nyg{\aa}rd}}, \bibinfo {author} {\bibfnamefont {K.~D.}\
  \bibnamefont {Petersson}}, \ and\ \bibinfo {author} {\bibfnamefont {C.~M.}\
  \bibnamefont {Marcus}},\ }\href@noop {} {\bibfield  {journal} {\bibinfo
  {journal} {arXiv preprint arXiv:1703.05643}\ } (\bibinfo {year}
  {2017})}\BibitemShut {NoStop}%
\bibitem [{\citenamefont {Casparis}\ \emph {et~al.}(2016)\citenamefont
  {Casparis}, \citenamefont {Larsen}, \citenamefont {Olsen}, \citenamefont
  {Kuemmeth}, \citenamefont {Krogstrup}, \citenamefont {Nyg{\aa}rd},
  \citenamefont {Petersson},\ and\ \citenamefont {Marcus}}]{casparis_2016}%
  \BibitemOpen
  \bibfield  {author} {\bibinfo {author} {\bibfnamefont {L.}~\bibnamefont
  {Casparis}}, \bibinfo {author} {\bibfnamefont {T.~W.}\ \bibnamefont
  {Larsen}}, \bibinfo {author} {\bibfnamefont {M.~S.}\ \bibnamefont {Olsen}},
  \bibinfo {author} {\bibfnamefont {F.}~\bibnamefont {Kuemmeth}}, \bibinfo
  {author} {\bibfnamefont {P.}~\bibnamefont {Krogstrup}}, \bibinfo {author}
  {\bibfnamefont {J.}~\bibnamefont {Nyg{\aa}rd}}, \bibinfo {author}
  {\bibfnamefont {K.~D.}\ \bibnamefont {Petersson}}, \ and\ \bibinfo {author}
  {\bibfnamefont {C.~M.}\ \bibnamefont {Marcus}},\ }\href@noop {} {\bibfield
  {journal} {\bibinfo  {journal} {Phys. Rev. Lett.}\ }\textbf {\bibinfo
  {volume} {116}},\ \bibinfo {pages} {150505} (\bibinfo {year}
  {2016})}\BibitemShut {NoStop}%
\bibitem [{sup()}]{sup}%
  \BibitemOpen
  \href@noop {} {}\bibinfo {note} {See the Supplemental Material at [URL to be
  inserted by publisher] for further details of the experimental
  setup.}\BibitemShut {Stop}%
\bibitem [{\citenamefont {Krogstrup}\ \emph {et~al.}(2015)\citenamefont
  {Krogstrup}, \citenamefont {Ziino}, \citenamefont {Chang}, \citenamefont
  {Albrecht}, \citenamefont {Madsen}, \citenamefont {Johnson}, \citenamefont
  {Nyg{\aa}rd}, \citenamefont {Marcus},\ and\ \citenamefont
  {Jespersen}}]{krogstrup_2015}%
  \BibitemOpen
  \bibfield  {author} {\bibinfo {author} {\bibfnamefont {P.}~\bibnamefont
  {Krogstrup}}, \bibinfo {author} {\bibfnamefont {N.~L.~B.}\ \bibnamefont
  {Ziino}}, \bibinfo {author} {\bibfnamefont {W.}~\bibnamefont {Chang}},
  \bibinfo {author} {\bibfnamefont {S.~M.}\ \bibnamefont {Albrecht}}, \bibinfo
  {author} {\bibfnamefont {M.~H.}\ \bibnamefont {Madsen}}, \bibinfo {author}
  {\bibfnamefont {E.}~\bibnamefont {Johnson}}, \bibinfo {author} {\bibfnamefont
  {J.}~\bibnamefont {Nyg{\aa}rd}}, \bibinfo {author} {\bibfnamefont {C.~M.}\
  \bibnamefont {Marcus}}, \ and\ \bibinfo {author} {\bibfnamefont {T.~S.}\
  \bibnamefont {Jespersen}},\ }\href@noop {} {\bibfield  {journal} {\bibinfo
  {journal} {Nat. Mater.}\ }\textbf {\bibinfo {volume} {14}},\ \bibinfo {pages}
  {400} (\bibinfo {year} {2015})}\BibitemShut {NoStop}%
\bibitem [{rot()}]{rotations}%
  \BibitemOpen
  \href@noop {} {}\bibinfo {note} {Rotations $R_{I}(\theta)=e^{\pm i\sigma_I
  \theta/2}$ ($I={X,Y,Z}$) are abbreviated in the text with $I$ for
  $\theta=\pi$ and $I/2$ for $\theta=\pi/2$.}\BibitemShut {Stop}%
\bibitem [{\citenamefont {Barends}\ \emph {et~al.}(2013)\citenamefont
  {Barends}, \citenamefont {Kelly}, \citenamefont {Megrant}, \citenamefont
  {Sank}, \citenamefont {Jeffrey}, \citenamefont {Chen}, \citenamefont {Yin},
  \citenamefont {Chiaro}, \citenamefont {Mutus}, \citenamefont {Neill},
  \citenamefont {O'Malley}, \citenamefont {Roushan}, \citenamefont {Wenner},
  \citenamefont {White}, \citenamefont {Cleland},\ and\ \citenamefont
  {Martinis}}]{barends_2013}%
  \BibitemOpen
  \bibfield  {author} {\bibinfo {author} {\bibfnamefont {R.}~\bibnamefont
  {Barends}}, \bibinfo {author} {\bibfnamefont {J.}~\bibnamefont {Kelly}},
  \bibinfo {author} {\bibfnamefont {A.}~\bibnamefont {Megrant}}, \bibinfo
  {author} {\bibfnamefont {D.}~\bibnamefont {Sank}}, \bibinfo {author}
  {\bibfnamefont {E.}~\bibnamefont {Jeffrey}}, \bibinfo {author} {\bibfnamefont
  {Y.}~\bibnamefont {Chen}}, \bibinfo {author} {\bibfnamefont {Y.}~\bibnamefont
  {Yin}}, \bibinfo {author} {\bibfnamefont {B.}~\bibnamefont {Chiaro}},
  \bibinfo {author} {\bibfnamefont {J.}~\bibnamefont {Mutus}}, \bibinfo
  {author} {\bibfnamefont {C.}~\bibnamefont {Neill}}, \bibinfo {author}
  {\bibfnamefont {P.}~\bibnamefont {O'Malley}}, \bibinfo {author}
  {\bibfnamefont {P.}~\bibnamefont {Roushan}}, \bibinfo {author} {\bibfnamefont
  {J.}~\bibnamefont {Wenner}}, \bibinfo {author} {\bibfnamefont {T.~C.}\
  \bibnamefont {White}}, \bibinfo {author} {\bibfnamefont {A.~N.}\ \bibnamefont
  {Cleland}}, \ and\ \bibinfo {author} {\bibfnamefont {J.~M.}\ \bibnamefont
  {Martinis}},\ }\href@noop {} {\bibfield  {journal} {\bibinfo  {journal}
  {Phys. Rev. Lett.}\ }\textbf {\bibinfo {volume} {111}},\ \bibinfo {pages}
  {080502} (\bibinfo {year} {2013})}\BibitemShut {NoStop}%
\bibitem [{\citenamefont {Macklin}\ \emph {et~al.}(2015)\citenamefont
  {Macklin}, \citenamefont {O'Brien}, \citenamefont {Hover}, \citenamefont
  {Schwartz}, \citenamefont {Bolkhovsky}, \citenamefont {Zhang}, \citenamefont
  {Oliver},\ and\ \citenamefont {Siddiqi}}]{macklin_2015}%
  \BibitemOpen
  \bibfield  {author} {\bibinfo {author} {\bibfnamefont {C.}~\bibnamefont
  {Macklin}}, \bibinfo {author} {\bibfnamefont {K.}~\bibnamefont {O'Brien}},
  \bibinfo {author} {\bibfnamefont {D.}~\bibnamefont {Hover}}, \bibinfo
  {author} {\bibfnamefont {M.~E.}\ \bibnamefont {Schwartz}}, \bibinfo {author}
  {\bibfnamefont {V.}~\bibnamefont {Bolkhovsky}}, \bibinfo {author}
  {\bibfnamefont {X.}~\bibnamefont {Zhang}}, \bibinfo {author} {\bibfnamefont
  {W.~D.}\ \bibnamefont {Oliver}}, \ and\ \bibinfo {author} {\bibfnamefont
  {I.}~\bibnamefont {Siddiqi}},\ }\href@noop {} {\bibfield  {journal} {\bibinfo
   {journal} {Science}\ }\textbf {\bibinfo {volume} {350}},\ \bibinfo {pages}
  {307} (\bibinfo {year} {2015})}\BibitemShut {NoStop}%
\bibitem [{\citenamefont {Chen}\ \emph
  {et~al.}(2014{\natexlab{b}})\citenamefont {Chen}, \citenamefont {Megrant},
  \citenamefont {Kelly}, \citenamefont {Barends}, \citenamefont {Bochmann},
  \citenamefont {Chen}, \citenamefont {Chiaro}, \citenamefont {Dunsworth},
  \citenamefont {Jeffrey}, \citenamefont {Mutus}, \citenamefont {O'Malley},
  \citenamefont {Neill}, \citenamefont {Roushan}, \citenamefont {Sank},
  \citenamefont {Vainsencher}, \citenamefont {Wenner}, \citenamefont {White},
  \citenamefont {Cleland},\ and\ \citenamefont {Martinis}}]{chen_air_2014}%
  \BibitemOpen
  \bibfield  {author} {\bibinfo {author} {\bibfnamefont {Z.}~\bibnamefont
  {Chen}}, \bibinfo {author} {\bibfnamefont {A.}~\bibnamefont {Megrant}},
  \bibinfo {author} {\bibfnamefont {J.}~\bibnamefont {Kelly}}, \bibinfo
  {author} {\bibfnamefont {R.}~\bibnamefont {Barends}}, \bibinfo {author}
  {\bibfnamefont {J.}~\bibnamefont {Bochmann}}, \bibinfo {author}
  {\bibfnamefont {Y.}~\bibnamefont {Chen}}, \bibinfo {author} {\bibfnamefont
  {B.}~\bibnamefont {Chiaro}}, \bibinfo {author} {\bibfnamefont
  {A.}~\bibnamefont {Dunsworth}}, \bibinfo {author} {\bibfnamefont
  {E.}~\bibnamefont {Jeffrey}}, \bibinfo {author} {\bibfnamefont {J.~Y.}\
  \bibnamefont {Mutus}}, \bibinfo {author} {\bibfnamefont {P.~J.~J.}\
  \bibnamefont {O'Malley}}, \bibinfo {author} {\bibfnamefont {C.}~\bibnamefont
  {Neill}}, \bibinfo {author} {\bibfnamefont {P.}~\bibnamefont {Roushan}},
  \bibinfo {author} {\bibfnamefont {D.}~\bibnamefont {Sank}}, \bibinfo {author}
  {\bibfnamefont {A.}~\bibnamefont {Vainsencher}}, \bibinfo {author}
  {\bibfnamefont {J.}~\bibnamefont {Wenner}}, \bibinfo {author} {\bibfnamefont
  {T.~C.}\ \bibnamefont {White}}, \bibinfo {author} {\bibfnamefont {A.~N.}\
  \bibnamefont {Cleland}}, \ and\ \bibinfo {author} {\bibfnamefont {J.~M.}\
  \bibnamefont {Martinis}},\ }\href@noop {} {\bibfield  {journal} {\bibinfo
  {journal} {Appl. Phys. Lett.}\ }\textbf {\bibinfo {volume} {104}},\ \bibinfo
  {pages} {052602} (\bibinfo {year} {2014}{\natexlab{b}})}\BibitemShut
  {NoStop}%
\bibitem [{Lan()}]{LandauZener}%
  \BibitemOpen
  \href@noop {} {}\bibinfo {note} {The level velocity depends on the slope of
  the qubit spectrum and rise time of the pulse detuning the qubit. We note
  that for the data in Fig. 4, the Q2 spectrum approaches a maximum at higher
  pulse amplitudes, resulting in prolonged swap oscillations.}\BibitemShut
  {Stop}%
\bibitem [{\citenamefont {Shim}\ and\ \citenamefont {Tahan}(2016)}]{shim_2016}%
  \BibitemOpen
  \bibfield  {author} {\bibinfo {author} {\bibfnamefont {Y.-P.}\ \bibnamefont
  {Shim}}\ and\ \bibinfo {author} {\bibfnamefont {C.}~\bibnamefont {Tahan}},\
  }\href {\doibase 10.1038/ncomms11059} {\bibfield  {journal} {\bibinfo
  {journal} {Nat. Commun.}\ }\textbf {\bibinfo {volume} {7}},\ \bibinfo {pages}
  {11059} (\bibinfo {year} {2016})}\BibitemShut {NoStop}%
\bibitem [{\citenamefont {Patel}\ \emph {et~al.}(2017)\citenamefont {Patel},
  \citenamefont {Pechenezhskiy}, \citenamefont {Plourde}, \citenamefont
  {Vavilov},\ and\ \citenamefont {McDermott}}]{patel_2016}%
  \BibitemOpen
  \bibfield  {author} {\bibinfo {author} {\bibfnamefont {U.}~\bibnamefont
  {Patel}}, \bibinfo {author} {\bibfnamefont {I.~V.}\ \bibnamefont
  {Pechenezhskiy}}, \bibinfo {author} {\bibfnamefont {B.~L.~T.}\ \bibnamefont
  {Plourde}}, \bibinfo {author} {\bibfnamefont {M.~G.}\ \bibnamefont
  {Vavilov}}, \ and\ \bibinfo {author} {\bibfnamefont {R.}~\bibnamefont
  {McDermott}},\ }\href {\doibase 10.1103/PhysRevB.96.220501} {\bibfield
  {journal} {\bibinfo  {journal} {Phys. Rev. B}\ }\textbf {\bibinfo {volume}
  {96}},\ \bibinfo {pages} {220501} (\bibinfo {year} {2017})}\BibitemShut
  {NoStop}%
\bibitem [{\citenamefont {Jochym-O'Connor}\ and\ \citenamefont
  {Bartlett}(2016)}]{bartlett_2016}%
  \BibitemOpen
  \bibfield  {author} {\bibinfo {author} {\bibfnamefont {T.}~\bibnamefont
  {Jochym-O'Connor}}\ and\ \bibinfo {author} {\bibfnamefont {S.~D.}\
  \bibnamefont {Bartlett}},\ }\href@noop {} {\bibfield  {journal} {\bibinfo
  {journal} {Phys. Rev. A}\ }\textbf {\bibinfo {volume} {93}},\ \bibinfo
  {pages} {022323} (\bibinfo {year} {2016})}\BibitemShut {NoStop}%
\bibitem [{\citenamefont {Buluta}\ and\ \citenamefont
  {Nori}(2009)}]{buluta_2009}%
  \BibitemOpen
  \bibfield  {author} {\bibinfo {author} {\bibfnamefont {I.}~\bibnamefont
  {Buluta}}\ and\ \bibinfo {author} {\bibfnamefont {F.}~\bibnamefont {Nori}},\
  }\href@noop {} {\bibfield  {journal} {\bibinfo  {journal} {Science}\ }\textbf
  {\bibinfo {volume} {326}},\ \bibinfo {pages} {108} (\bibinfo {year}
  {2009})}\BibitemShut {NoStop}%
\end{thebibliography}
%

\end{document}